\documentclass[useAMS,usenatbib]{mn2e}
\usepackage{tabularx}
\usepackage{xspace}
\usepackage{graphicx}
\newcommand{\ignore}[1]{}
\providecommand{\ao}{}

\renewcommand{\ao}{adaptive optics (AO)\renewcommand{\ao}{AO\xspace}\renewcommand{\Ao}{AO\xspace}\xspace}
\newcommand{\Ao}{Adaptive optics (AO)\renewcommand{\ao}{AO\xspace}\renewcommand{\Ao}{AO\xspace}\xspace}
\newcommand{\wfs}{wavefront sensor (WFS)\renewcommand{\wfs}{WFS\xspace}\renewcommand{\wfss}{WFSs\xspace}\xspace}
\newcommand{\wfss}{wavefront sensors (WFSs)\renewcommand{\wfs}{WFS\xspace}\renewcommand{\wfss}{WFSs\xspace}\xspace}
\newcommand{\shwfs}{Shack-Hartmann \wfs (SHWFS)\renewcommand{\shwfs}{SHWFS\xspace}\xspace}
\newcommand{\dm}{deformable mirror (DM)\renewcommand{\dm}{DM\xspace}\renewcommand{\dms}{DMs\xspace}\renewcommand{\Dms}{DMs\xspace}\renewcommand{\Dm}{DM\xspace}\xspace}
\newcommand{\dms}{deformable mirrors (DMs)\renewcommand{\dm}{DM\xspace}\renewcommand{\dms}{DMs\xspace}\renewcommand{\Dms}{DMs\xspace}\renewcommand{\Dm}{DM\xspace}\xspace}
\newcommand{\Dms}{Deformable mirrors (DMs)\renewcommand{\dm}{DM\xspace}\renewcommand{\dms}{DMs\xspace}\renewcommand{\Dms}{DMs\xspace}\renewcommand{\Dm}{DM\xspace}\xspace}
\newcommand{\Dm}{Deformable mirror (DM)\renewcommand{\dm}{DM\xspace}\renewcommand{\dms}{DMs\xspace}\renewcommand{\Dms}{DMs\xspace}\renewcommand{\Dm}{DM\xspace}\xspace}

\newcommand{\lqg}{linear-quadratic-gaussian (LQG)\renewcommand{\lqg}{LQG\xspace}\xspace}
\newcommand{\shs}{Shack-Hartmann sensor (SHS)\renewcommand{\shs}{SHS\xspace}\renewcommand{\shss}{SHSs\xspace}\xspace}
\newcommand{\shss}{Shack-Hartmann sensors (SHSs)\renewcommand{\shs}{SHS\xspace}\renewcommand{\shss}{SHSs\xspace}\xspace}
\newcommand{\lgs}{laser guide star (LGS)\renewcommand{\lgs}{LGS\xspace}\renewcommand{\Lgs}{LGS\xspace}\renewcommand{\lgss}{LGSs\xspace}\xspace}
\newcommand{\lgss}{laser guide stars (LGSs)\renewcommand{\lgs}{LGS\xspace}\renewcommand{\Lgs}{LGS\xspace}\renewcommand{\lgss}{LGSs\xspace}\xspace}
\newcommand{\Lgs}{Laser guide star (LGS)\renewcommand{\lgs}{LGS\xspace}\renewcommand{\Lgs}{LGS\xspace}\renewcommand{\lgss}{LGSs\xspace}\xspace}

\newcommand{\Ngs}{Natural guide star (NGS)\renewcommand{\ngs}{NGS\xspace}\renewcommand{\Ngs}{NGS\xspace}\renewcommand{\ngss}{NGSs\xspace}\xspace}
\newcommand{\ngs}{natural guide star (NGS)\renewcommand{\ngs}{NGS\xspace}\renewcommand{\Ngs}{NGS\xspace}\renewcommand{\ngss}{NGSs\xspace}\xspace}
\newcommand{\ngss}{natural guide stars (NGSs)\renewcommand{\ngs}{NGS\xspace}\renewcommand{\Ngs}{NGS\xspace}\renewcommand{\ngss}{NGSs\xspace}\xspace}
\newcommand{\mems}{Micro-Electro-Mechanical Systems (MEMS)\renewcommand{\mems}{MEMS\xspace}\xspace}
\newcommand{\snr}{signal to noise ratio (SNR)\renewcommand{\snr}{SNR\xspace}\xspace}
\newcommand{\Moao}{Multi-object \ao (MOAO)\renewcommand{\moao}{MOAO\xspace}\renewcommand{\Moao}{MOAO\xspace}\xspace}
\newcommand{\moao}{multi-object \ao (MOAO)\renewcommand{\moao}{MOAO\xspace}\renewcommand{\Moao}{MOAO\xspace}\xspace}
\newcommand{\mcao}{multi-conjugate adaptive optics (MCAO)\renewcommand{\mcao}{MCAO\xspace}\xspace}
\newcommand{\ltao}{laser tomographic \ao (LTAO)\renewcommand{\ltao}{LTAO\xspace}\xspace}
\newcommand{\cpu}{central processing unit (CPU)\renewcommand{\cpu}{CPU\xspace}\renewcommand{\cpus}{CPUs\xspace}\xspace}
\newcommand{\cpus}{central processing units (CPUs)\renewcommand{\cpu}{CPU\xspace}\renewcommand{\cpus}{CPUs\xspace}\xspace}
\newcommand{\psf}{point spread function (PSF)\renewcommand{\psf}{PSF\xspace}\renewcommand{\psfs}{PSFs\xspace}\renewcommand{\Psf}{PSF\xspace}\xspace}
\newcommand{\psfs}{point spread functions (PSFs)\renewcommand{\psf}{PSF\xspace}\renewcommand{\psfs}{PSFs\xspace}\renewcommand{\Psf}{PSF\xspace}\xspace}
\newcommand{\Psf}{Point spread function (PSF)\renewcommand{\psf}{PSF\xspace}\renewcommand{\psfs}{PSFs\xspace}\renewcommand{\Psf}{PSF\xspace}\xspace}
\newcommand{\fpga}{field programmable gate array (FPGA)\renewcommand{\fpga}{FPGA\xspace}\renewcommand{\fpgas}{FPGAs\xspace}\xspace}
\newcommand{\fpgas}{field programmable gate arrays (FPGAs)\renewcommand{\fpga}{FPGA\xspace}\renewcommand{\fpgas}{FPGAs\xspace}\xspace}
\newcommand{\sor}{successive over-relaxation (SOR)\renewcommand{\sor}{SOR\xspace}\xspace}
\newcommand{\fdpcg}{Fourier domain pre-conditioned gradient (FDPCG)\renewcommand{\fdpcg}{FDPCG\xspace}\xspace}
\newcommand{\map}{maximum a-posteriori (MAP)\renewcommand{\map}{MAP\xspace}\xspace}

\newcommand{\elt}{Extremely Large Telescope (ELT)\renewcommand{\elt}{ELT\xspace}\renewcommand{\elts}{ELTs\xspace}\renewcommand{\eelt}{European ELT (E-ELT)\renewcommand{\eelt}{E-ELT\xspace}\xspace}\xspace}

\newcommand{\elts}{Extremely Large Telescopes (ELTs)\renewcommand{\elt}{ELT\xspace}\renewcommand{\elts}{ELTs\xspace}\renewcommand{\eelt}{European ELT (E-ELT)\renewcommand{\eelt}{E-ELT\xspace}\xspace}\xspace}

\newcommand{\eelt}{European Extremely Large Telescope (E-ELT)\renewcommand{\eelt}{E-ELT\xspace}\renewcommand{\elt}{ELT\xspace}\renewcommand{\elts}{ELTs\xspace}\xspace}

\newcommand{\dugall}{Durham University generalised adaptive optics laser laboratory (DUGALL)\renewcommand{\dugall}{DUGALL\xspace}\xspace}
\newcommand{\fwhm}{full-width at half-maximum (FWHM)\renewcommand{\fwhm}{FWHM\xspace}\xspace}
\newcommand{\wht}{William Herschel Telescope (WHT)\renewcommand{\wht}{WHT\xspace}\xspace}
\newcommand{\emccd}{electron multiplying CCD (EMCCD)\renewcommand{\emccd}{EMCCD\xspace}\renewcommand{\emccds}{EMCCDs\xspace}\xspace}
\newcommand{\emccds}{electron multiplying CCDs (EMCCDs)\renewcommand{\emccd}{EMCCD\xspace}\renewcommand{\emccds}{EMCCDs\xspace}\xspace}
\newcommand{\dasp}{Durham \ao simulation platform (DASP)\renewcommand{\dasp}{DASP\xspace}\renewcommand{\thedasp}{DASP\xspace}\renewcommand{\Thedasp}{DASP\xspace}\xspace}
\newcommand{\thedasp}{the Durham \ao simulation platform (DASP)\renewcommand{\dasp}{DASP\xspace}\renewcommand{\thedasp}{DASP\xspace}\renewcommand{\Thedasp}{DASP\xspace}\xspace}
\newcommand{\Thedasp}{The Durham \ao simulation platform (DASP)\renewcommand{\dasp}{DASP\xspace}\renewcommand{\thedasp}{DASP\xspace}\renewcommand{\Thedasp}{DASP\xspace}\xspace}
\newcommand{\mpi}{Message Passing Interface (MPI)\renewcommand{\mpi}{MPI\xspace}\xspace}
\newcommand{\smp}{symmetric multi-processing (SMP)\renewcommand{\smp}{SMP\xspace}\xspace}
\newcommand{\svd}{singular value decomposition (SVD)\renewcommand{\svd}{SVD\xspace}\xspace}
\newcommand{\gpu}{graphics processing unit (GPU)\renewcommand{\gpu}{GPU\xspace}\renewcommand{\gpus}{GPUs\xspace}\xspace}
\newcommand{\gpus}{graphics processing units (GPUs)\renewcommand{\gpu}{GPU\xspace}\renewcommand{\gpus}{GPUs\xspace}\xspace}
\newcommand{\fft}{fast Fourier transform (FFT)\renewcommand{\fft}{FFT\xspace}\xspace}
\newcommand{\ifu}{integral field unit (IFU)\renewcommand{\ifu}{IFU\xspace}\xspace}
\newcommand{\darc}{the Durham \ao real-time controller (DARC)\renewcommand{\darc}{DARC\xspace}\renewcommand{\Darc}{DARC\xspace}\xspace}
\newcommand{\Darc}{The Durham \ao real-time controller (DARC)\renewcommand{\darc}{DARC\xspace}\renewcommand{\Darc}{DARC\xspace}\xspace}
\newcommand{\cots}{commercial off-the-shelf (COTS)\renewcommand{\cots}{COTS\xspace}\xspace}
\newcommand{\rtcp}{real-time control pipeline (RTCP)\renewcommand{\rtcp}{RTCP\xspace}\xspace}
\newcommand{\rms}{root-mean-square (RMS)\renewcommand{\rms}{RMS\xspace}\xspace}
\newcommand{\sFPDP}{serial Front Panel Data Port (sFPDP)\renewcommand{\sFPDP}{sFPDP\xspace}\xspace}
\newcommand{\wpu}{wavefront processing unit (WPU)\renewcommand{\wpu}{WPU\xspace}\xspace}

\newcommand{\rtcs}{real-time control system (RTCS)\renewcommand{\rtcs}{RTCS\xspace}\renewcommand{\rtcss}{RTCSs\xspace}\xspace}
\newcommand{\rtcss}{real-time control systems (RTCSs)\renewcommand{\rtcs}{RTCS\xspace}\renewcommand{\rtcss}{RTCSs\xspace}\xspace}
\newcommand{\eso}{European Southern Observatory (ESO)\renewcommand{\eso}{ESO\xspace}\renewcommand{\theeso}{ESO\xspace}\xspace}
\newcommand{\theeso}{\renewcommand{\theeso}{ESO\xspace}the \eso}
\newcommand{\scao}{single conjugate \ao (SCAO)\renewcommand{\scao}{SCAO\xspace}\renewcommand{\Scao}{SCAO\xspace}\xspace}
\newcommand{\Scao}{Single conjugate \ao (SCAO)\renewcommand{\scao}{SCAO\xspace}\renewcommand{\Scao}{SCAO\xspace}\xspace}
\newcommand{\glao}{ground layer \ao (GLAO)\renewcommand{\glao}{GLAO\xspace}\xspace}
\newcommand{\eagle}{ELT Adaptive optics for GaLaxy Evolution (EAGLE)\renewcommand{\eagle}{EAGLE\xspace}\xspace}
\newcommand{\maory}{multi-conjugate \ao relay for the \eelt (MAORY)\renewcommand{\maory}{MAORY\xspace}\xspace}
\newcommand{\muse}{Multi Unit Spectroscopic Explorer (MUSE)\renewcommand{\muse}{MUSE\xspace}\xspace}
\newcommand{\vlt}{Very Large Telescope (VLT)\renewcommand{\vlt}{VLT\xspace}\xspace}

\newcommand{\eapd}{electron avalanche photodiode\renewcommand{\eapd}{eAPD\xspace}\xspace}
\newcommand{\tmt}{Thirty Metre Telescope (TMT)\renewcommand{\tmt}{TMT\xspace}\xspace}
\newcommand{\lbt}{Large Binocular Telescope (LBT)\renewcommand{\lbt}{LBT\xspace}\xspace}
\newcommand{\xao}{eXtreme \ao (XAO)\renewcommand{\xao}{XAO\xspace}\xspace}

\newcommand{\vla}{Very Large Array (VLA)\renewcommand{\vla}{VLA\xspace}\xspace}
\newcommand{\jwst}{{\em James Webb Space Telescope} \citep[JWST,][]{jwst}\renewcommand{\jwst}{{\em JWST}\xspace}\xspace}
\newcommand{\hst}{{\em Hubble Space Telescope (HST)}\renewcommand{\hst}{{\em HST}\xspace}\xspace}
\newcommand{\ifss}{integral-field spectrographs (IFSs)\renewcommand{\ifss}{IFSs\xspace}\renewcommand{\ifs}{IFS\xspace}\xspace}
\newcommand{\ifs}{integral-field spectrograph (IFS)\renewcommand{\ifss}{IFSs\xspace}\renewcommand{\ifs}{IFS\xspace}\xspace}
\newcommand{\ifus}{integral field units (IFUs)\renewcommand{\ifus}{IFUs\xspace}\xspace}
\newcommand{\mos}{multi-object spectrograph (MOS)\renewcommand{\mos}{MOS\xspace}\xspace}
\newcommand{\goodss}{Great Observatories Origins Deep Survey (GOODS)-S\renewcommand{\goodss}{GOODS-S\xspace}\xspace}
\newcommand{\goods}{Great Observatories Origins Deep Survey (GOODS)\renewcommand{\goods}{GOODS\xspace}\xspace}
\newcommand{\cmos}{complimentary metal-oxide semiconductor (CMOS)\renewcommand{\cmos}{CMOS\xspace}\xspace}
\newcommand{\scmos}{scientific CMOS (sCMOS)\renewcommand{\scmos}{sCMOS\xspace}\xspace}
\newcommand{\aof}{Adaptive Optics Facility (AOF)\renewcommand{\aof}{AOF\xspace}\xspace}
\newcommand{\dsp}{digital signal processor (DSP)\renewcommand{\dsp}{DSP\xspace}\renewcommand{\dsps}{DSPs\xspace}\xspace}
\newcommand{\dsps}{digital signal processors (DSPs)\renewcommand{\dsp}{DSP\xspace}\renewcommand{\dsps}{DSPs\xspace}\xspace}
\newcommand{\capi}{Coherent Accelerator Processor Interface (CAPI)\renewcommand{\capi}{CAPI\xspace}\xspace}
\newcommand{\qe}{quantum efficiency (QE)\renewcommand{\qe}{QE\xspace}\xspace}
\newcommand{\numa}{non-uniform memory access (NUMA)\renewcommand{\numa}{NUMA\xspace}\xspace}
\newcommand{\uav}{unmanned aerial vehicle (UAV)\renewcommand{\uav}{UAV\xspace}\renewcommand{\uavs}{UAVs\xspace}\xspace}
\newcommand{\uavs}{unmanned aerial vehicles (UAVs)\renewcommand{\uav}{UAV\xspace}\renewcommand{\uavs}{UAVs\xspace}\xspace}

\newcommand{\ncpa}{non-common path aberration (NCPA)\renewcommand{\ncpa}{NCPA\xspace}\renewcommand{\ncpas}{NCPAs\xspace}\xspace}
\newcommand{\ncpas}{non-common path aberrations (NCPA)\renewcommand{\ncpa}{NCPA\xspace}\renewcommand{\ncpas}{NCPAs\xspace}\xspace}
\newcommand{\sdk}{software developers kit (SDK)\renewcommand{\sdk}{SDK\xspace}\renewcommand{\sdks}{SDKs\xspace}\xspace}
\newcommand{\sdks}{software developers kits (SDKs)\renewcommand{\sdk}{SDK\xspace}\renewcommand{\sdks}{SDKs\xspace}\xspace}
\newcommand{\dac}{digital to analogue converter (DAC)\renewcommand{\dac}{DAC\xspace}\xspace}
\newcommand{\nda}{non-disclosure agreement (NDA)\renewcommand{\nda}{NDA\xspace}\xspace}
\newcommand{\polc}{pseudo-open-loop control (POLC)\renewcommand{\polc}{POLC\xspace}\xspace}
\newcommand{\udp}{User Datagram Protocol (UDP)\renewcommand{\udp}{UDP\xspace}\xspace}
\newcommand{\ags}{artificial guide star (AGS)\renewcommand{\ags}{AGS\xspace}\xspace}

\usepackage{amssymb}

\title[E-ELT MOAO performance modelling]{Monte-Carlo
  modelling of multi-object adaptive optics performance on the
  European Extremely Large Telescope}

\author[A.\ G.\ Basden et al.]{A.\ G.\ Basden$^{1}$\thanks{E-mail:
    a.g.basden@durham.ac.uk (AGB)}, T.\ J.\ Morris$^1$\\
$^{1}$Department of Physics, South Road, Durham, DH1 3LE, UK}

\begin{document}
\maketitle

\begin{abstract}
The performance of a wide-field adaptive optics system depends on
input design parameters.  Here we investigate the performance of a
multi-object adaptive optics system design for the European Extremely
Large Telescope, using an end-to-end Monte-Carlo adaptive optics
simulation tool, DASP, with relevance for proposed instruments such as
MOSAIC.  We consider parameters such as the number of laser guide
stars, sodium layer depth, wavefront sensor pixel scale, actuator
pitch and natural guide star availability.  We provide potential areas
where costs savings can be made, and investigate trade-offs between
performance and cost, and provide solutions that would enable such an
instrument to be built with currently available technology.  Our key
recommendations include a trade-off for laser guide star wavefront
sensor pixel scale of about 0.7~arcseconds per pixel, and a field of
view of at least 7~arcseconds, that EMCCD
technology should be used for natural guide star wavefront sensors
even if reduced frame rate is necessary, and that sky coverage can be
improved by a slight reduction in natural guide star sub-aperture
count without significantly affecting tomographic performance.  We find that
adaptive optics correction can be maintained across a wide field of
view, up to 7~arcminutes in diameter.  We also recommend the use of at
least 4 laser guide stars, and include ground-layer and multi-object
adaptive optics performance estimates.
\end{abstract}


\section{Introduction}
\label{sect:intro}
The next generation of ground based astronomical telescopes will be
the  \elts \citep{eelt,tmt,gmt}, which are due to see first light
within the next decade.  All of these telescopes rely on \ao systems
\citep{adaptiveoptics} to provide compensation for the degrading
effects of atmospheric turbulence, thus allowing the scientific
goals of these facilities to be met.  Extensive simulation of \ao
systems is required during the instrument design phases, so that
predicted performance estimates can be made and design trade-offs
explored and \ao designs optimised.

High fidelity modelling of the performance of \ao and telescope
systems can be implemented using Monte-Carlo simulation, which
involves playing a time sequence of input atmospheric perturbations
through the \ao system and telescope models.  For \elt-scale instruments, these
simulations are computationally expensive.  Here, we report on an
exploration of the parameter space related to a \moao system design
study for the 39~m \eelt.  We use \thedasp \citep{basden5,basden11} to
perform this modelling.  Our models are based on a system with both
\lgss and \ngss (the number of which we explore), and we use a high
resolution atmospheric model \citep[as used in previous studies,
  e.g.\ ][]{basden21} which is stratified into 35
discrete layers of turbulence.  Our results have relevance for
proposed and future \moao instruments such as MOSAIC, and also more
generally for other wide-field \ao systems.

Within this study, we investigate factors such as the number of \lgss,
the elongation of \lgss as seen by the \wfss (due to the extent of the
mesosphere sodium layer depth),
the number, position and magnitude of \ngss within the field of view,
\dm requirements, detector requirements, \wfs sensitivity and field of
view and the effect of turbulence strength.  We explore \ao
performance across the field of view, and our default results are
presented on-axis, which we show to be pessimistic compared to most of
the rest of the field of view.  We also consider the performance
improvements achievable by operating the \lgs and \ngs \wfss at
different frame rates, and consider the use of currently available
commercial cameras as wavefront sensors.

The results that we present can be used to aid design decisions for
\elt instrumentation, and also to provide a benchmark for simulation
comparison.  These results are complementary to those from other
modelling tools for \elt \moao instrumentation (which for a single
on-axis channel can be viewed as \ltao), for example
\citet{miskaltao,2014SPIE.9148E..6FA}.

In \S2 we present the key parameters of our simulations, and details
of the parameter space that is explored, along with key algorithms.
In \S3 we present our resulting estimates of \ao system performance, and we conclude in
\S4.

\section{ELT-scale MOAO system modelling}
\begin{figure*}
  \includegraphics[width=\linewidth]{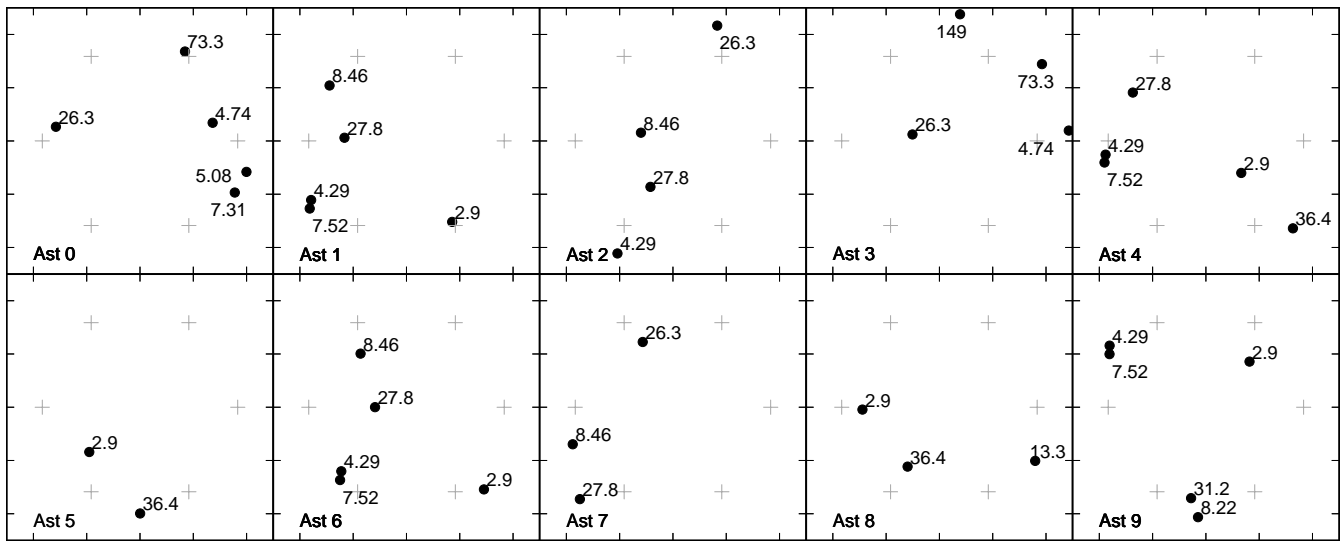}
  \caption{A figure showing the NGS asterisms, and LGS positions, that
    are used throughout this study.  The numbers next to the guide
    stars represent flux in photons per sub-aperture per frame at
    250~Hz.  The LGS positions are given by +.  Asterism number is
    denoted by ``Ast''.  The total field of view across each asterism
    is 10~arcminutes, and the axis tick marks are 2~arcminutes apart.}
  \label{fig:asterisms}
\end{figure*}

We base our modelling on \dasp, which has been cross checked and
verified against other \ao simulation codes, and which has a long
history of \ao system modelling.  In a previous study
\citep{basden17}, we explored many different \ngs asterisms available
within a cosmological field (which by definition are relatively free
of suitable bright guide stars), and the effect that these have on \ao
performance.  Here, we base this study on the same set of \ngss, as
shown in Fig.~\ref{fig:asterisms}.  Unless otherwise stated, we use
asterism 0 as the default case, as this provides a close-to-median
performance.  Conversion from guide star magnitude to detected \wfs
flux level is given by \citet{basden17}

The key \ao performance metric that we present here is ensquared
energy within a 150~mas box diameter at H-band (1650~nm wavelength).
The reason for this is that for a wide-field \moao system with limited
guide star numbers (both \ngs and \lgs), Strehl ratio is typically
low, and ensquared energy is therefore more sensitive to changes in
\ao performance
as parameter space is explored.  Additionally, \moao instruments are
typically coupled to spectrographs, and of relevance here is how much
light (energy) can be fed into an optical fibre.  However, we also
consider other performance metrics, and performance at other
wavelengths.  We ensure that the science \psfs are well averaged, and
typically integrate for 20~s of telescope time (5,000 iterations).
The uncertainties in our results due to Monte-Carlo randomness are
below 1\%, which we have verified using a suite of separate
Monte-Carlo instantiations.    

We ignore the tip-tilt signal from the \lgss (since this is generally
not known), and we use the \ngss for full high order correction, in
addition to tip-tilt correction.  Where \ngs flux is particularly low,
we investigate reducing the frame rate of these particular \wfss (to
increase the flux).  When doing this, wavefront reconstruction is
performed using the newest available measurement, at the rate of the
fastest wavefront sensor, i.e.\ a zero-order hold for the slower
\wfss.  We do not investigate more complicated algorithms.

The \eelt design has a ``deformable secondary'' \dm (actually the
fourth mirror in the optical train, M4).  Our simulations therefore
have a 2-\dm design, using this M4 \dm to perform a global ground
layer correction, and then individual \moao \dms to compensate along a
specific line of sight.  We assume that M4 is optically conjugated to
the ground layer, though in the \eelt design it is actually conjugated
at about 625~m.  However, a previous study has shown that this has a
negligible effect on \ao performance \citep{basden21}.

\subsection{Details of the simulation model}
The simulations presented here use a standard \eso turbulence profile
for the \eelt site \citep{35layer}, containing 35 turbulent layers
extending up to about 20~km.  The default outer scale is 25~m with a
13.5~cm Fried's parameter (at zenith).  Five variations of this
profile are studied, with the median seeing case, and one case for
each quartile.  Unless stated otherwise, we assume the median profile,
and we observe at 30$^\circ$ from zenith.  We assume an \ao system
update rate of 250~Hz, which is fairly typical for \moao systems
\citep{2014SPIE.9148E..1GL,canaryresultsshort,basden22short} and has
been used in previous studies of \elt \moao systems \citep{basden15}.
For wide-field \ao systems, \ao latency does not tend to dominate the
error budget, and therefore we do not present any investigations of
\ao frame rate here, other than reductions in the speed of faint \ngss
as mentioned previously.

We assume a primary mirror diameter (largest optical diameter) of
38.55~m, and the M4 \dm has $75\times75$ actuators in a square grid by
default.  We also compare performance with a hexagonal actuator
pattern.  We assume a telescope central obscuration diameter of 11~m,
and our models include the hexagonal edge pattern based on a primary
mirror design created from 798 segmented hexagonal mirrors, matching
the \eelt.
The telescope pupil function is modelled as direction dependent, with
vignetting by the central obscuration changing depending on line of
sight.  We also include telescope support structures (spiders) in this
pupil function, following models used by \citet{basden21}.

\subsubsection{Wavefront sensors}
In our default case, the wavefront sensors all have $74\times74$
sub-apertures.  We also consider \ao performance when the number of
\ngs \wfs sub-apertures are reduced (to increase individual sub-aperture flux). We
use up to 5 \ngss (at positions given by Fig.~\ref{fig:asterisms}),
and up to 6 \lgss (6 for the default case) equally spaced around a
circular asterism which has a default diameter of 7.3~arcmin.  The
\lgss are side-launched from four launch locations, spread equally
around the telescope, 22~m from the central axis.  Our default
simulation uses a sodium layer with a Gaussian profile and a \fwhm
depth of 10~km, centred at 90~km above the ground (with focal
anisoplanatism).  We assume that the \lgs \psfs have a 1~arcsecond
\fwhm across the spot profile, due to a combination of atmospheric
broadening and a finite width laser plume.  Combined with our default
\lgs pixel scale and sub-aperture size, this results in a small amount of spot truncation
with sub-apertures furthest from the laser launch aperture having flux
reduced by about 2.5\% due to truncation.  The default \lgs flux is
5000 photons per sub-aperture per frame (approximately 6 million
photons~m$^{-2}$~s$^{-1}$ on-sky with a 90\% telescope throughput and
an 85\% \wfs throughput), and we investigate other
signal levels.  This flux level is chosen based on measurements of
sodium layer return flux by the \eso Wendelstein \lgs unit
\citep{lgsflux} which returns between 5--21 million
photons~m$^{-2}$~s$^{-1}$.

Sub-apertures typically have $16\times16$ pixels, unless otherwise
stated.  For the \lgs \wfss, this helps to reduce the effects of spot
truncation.  For the \ngs \wfss, such a large field of view is not
strictly necessary, but is used for simplicity.  We do however
investigate smaller sub-apertures.  We include readout noise, with a
default value of 0.1~electrons per pixel, corresponding to \emccd
technology, and investigate performance with up to 3~electrons readout
noise, corresponding to equivalent levels experienced by \scmos
technology detectors due to the non-Gaussian distribution of readout
noise \citep{basden19}.  Photon shot noise is also included.  We apply
a threshold to the sub-apertures before slope calculation which has a
default value of three times the readout noise.

Our default wavefront sensor pixels scale is 0.7~arcseconds per pixel
for the \lgs \wfss, and 0.25~arcseconds per pixel for the \ngs \wfss.

Since instrumental and telescope optical throughputs, and precise
detector quantum efficiencies are not well known, we also investigate
\ao performance when \ngs flux is increased and reduced, to provide an
estimation for sensitivity to changes in detected flux.

\subsubsection{Wavefront reconstruction}
We perform tomographic wavefront reconstruction using a virtual \dm
approach.  First, wavefront sensor signals are reconstructed at
several discrete heights to give an estimate of the wavefront phase at
these locations.  Projection along a given line of sight then provides
the signal which is sent to the individual \moao \dms, which are
conjugated to the telesecope pupil.  We find that using 12 such
virtual \dms gives good performance, with little gained by using an
increased number (and using fewer leads to reduced performance).  The
virtual \dms are conjugated close to dominant atmospheric layers, and
the position of these is modified when using different atmospheric profiles.

The pitch of phase reconstruction is dependent on layer strength
rather than constant \citep{gavelDMFittingError}, to help reduce
computational load, though is fixed at the \lgs sub-aperture pitch for
the ground conjugate \dm.  We find that slightly improved performance
can be obtained using an increased number of phase reconstruction
points for non-ground conjugate \dms, though do not present this here.
Wavefront reconstruction is based on a minimum mean square error
(MMSE) algorithm \citep{map}.  We approximate wavefront phase
covariance using a Laplacian regularisation, and noise covariance is
approximated to a single value for each wavefront sensor, dependent on
flux and readout noise.  Although this is slightly sub-optimal, it
allows us to simplify our modelling.

\subsubsection{Deformable mirrors}
The \dms are modelled using a cubic spline interpolation function,
which uses given actuator heights and positions to compute a surface
map of the \dm.  The \moao \dms have $64\times64$ sub-apertures by
default, though we also investigate lower actuator counts, to match a
range of commercially available \dms.  A previous study has
investigated required stroke and \dm imperfections \citep{basden15}.

\section{Wide field of view MOAO performance estimation for the E-ELT}
When designing an \ao system, maximum performance is always
desirable.  However, budget limitations usually mean that design
trade-offs must be made.  Here, we present results from several
trade-off studies that will allow \elt \moao performance to be
optimised within a given budget.

\begin{figure*}
  \includegraphics[width=0.8\linewidth]{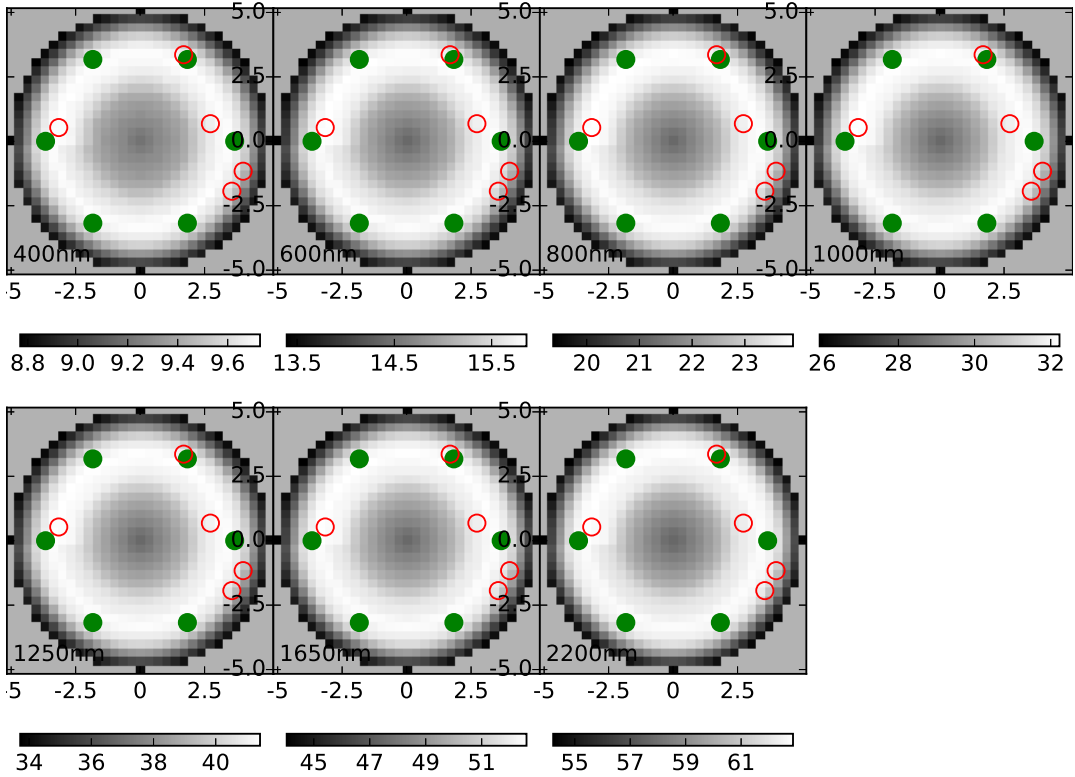}
  \caption{A figure showing predicted ensquared energy in a 150~mas
    box across a 10~arcminute field of view at the wavelengths given
    on the sub-figures.  The LGS positions are shown by filled green
    circles, and the NGS positions by unfilled red circles.}
  \label{fig:eemap}
\end{figure*}

Fig.~\ref{fig:eemap} shows predicted \ao performance (ensquared energy
within a 150~mas box size) across the 10 arcminute field of view for
our default simulation case, at multiple wavelength bands.  It can be
seen that within the \lgs asterism diameter, \ao correction is
relatively uniform, but that performance drops faster outside.

\subsection{The effect of LGS WFS pixel scale on AO performance}
\begin{figure}
  \includegraphics[width=\linewidth]{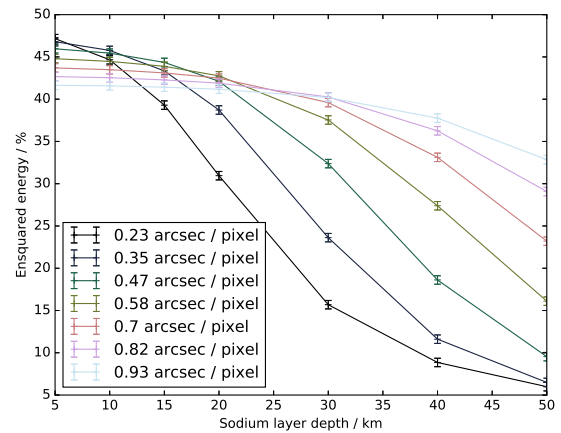}
  \caption{A figure showing on-axis \ao performance (H-band ensquared
    energy within a 150~mas box) as a function of sodium layer depth
    (full width half maximum) for different pixel scales as given by
    the legend.}
  \label{fig:pxlscale}
\end{figure}

The sensitivity of a \wfs is determined in part by its pixel scale.
For elongated \lgs spots, there is a trade-off between the number of
detector pixels per sub-aperture (each of which introduces noise), the
spot size (with elongation determined by off-axis distance and sodium
layer profile), the expected range of spot motion, and the number of pixels
over which flux is distributed.  Fig.~\ref{fig:pxlscale} shows
predicted on-axis \ao performance as a function of sodium layer depth
for different pixel scales (resulting in different elongation of
Shack-Hartmann spots).  In this case, each sub-aperture has
$16\times16$ pixels, and receives a mean flux of 5000 photons per frame.
It can be seen here that for larger sodium layer depths, it is
favourable to have larger pixel scales.  We have therefore selected
0.7~arcseconds per pixel as the default for this study, since
performance remains good up to depths of 30~km which is a likely upper
limit \citep{pfrommerdata}.  We note that a smaller pixel scale would lead to an
improvement of a few percent in ensquared energy when sodium layer
depth is small.  However, this gain quickly drops off as the depth
increases, and 0.7~arcseconds per pixel is a good compromise.  

\begin{figure}
\includegraphics[width=\linewidth]{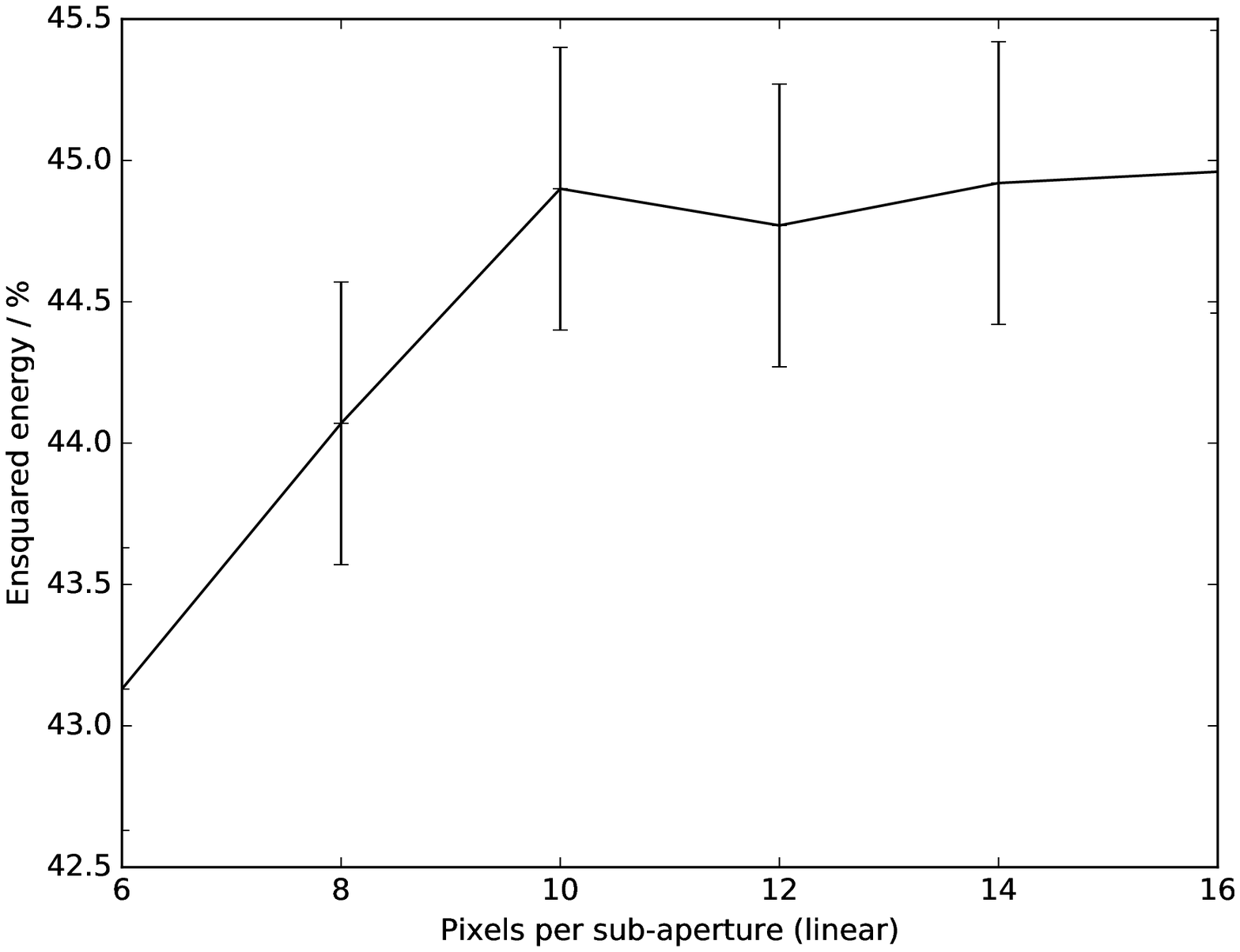}
\includegraphics[width=\linewidth]{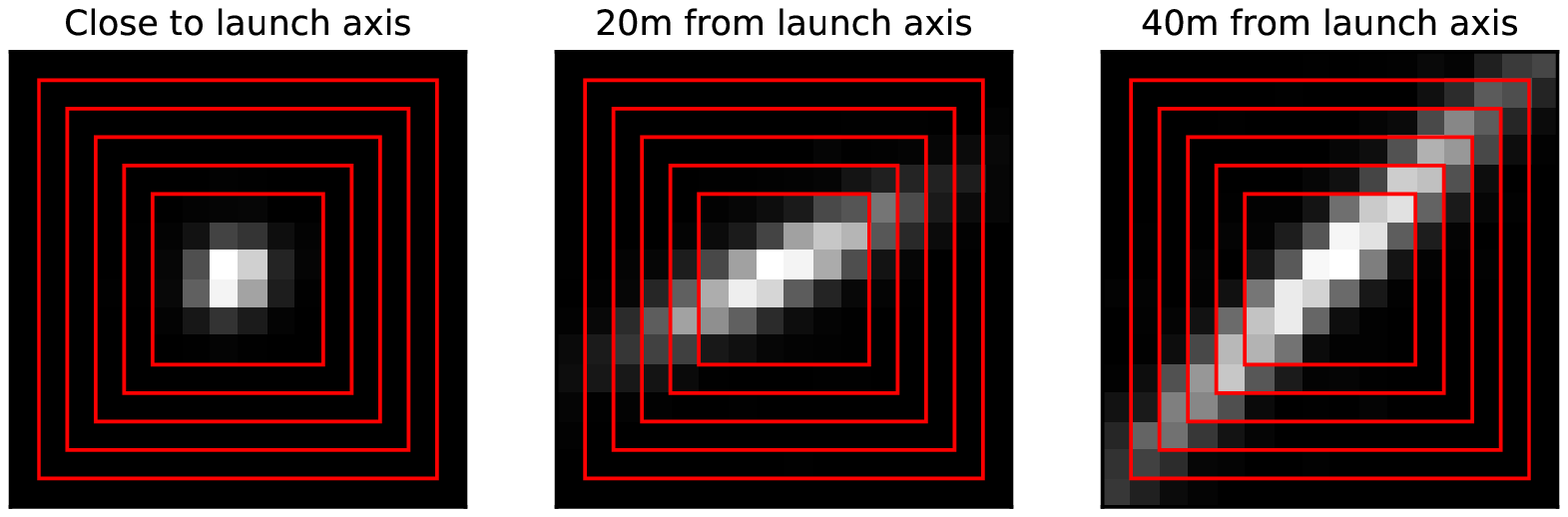}
\caption{(a) A figure showing AO performance (H-band ensquared energy
  in 150~mas) as a function of linear sub-aperture size in pixels.
  The pixel scale is kept constant at 0.7~arcsec per pixel.
  (b) Showing the LGS truncation for different sub-apertures at
  increasing distances from the LGS launch axis.  The boxes show
  sub-aperture sizes from $6\times6$ to $16\times16$ pixels, and the
  sodium layer depth FWHM is 10~km}
\label{fig:lgsncen}
  \end{figure}

We also investigate the effect of a reduction in number of pixels per
\lgs sub-aperture.  This is
of interest in instrument designs because a requirement for fewer
pixels equates to smaller detectors, increasing the likelihood of
commercial availability.  Additionally, fewer pixels have reduced
readout noise, increasing performance, but also contribute to greater
truncation of elongated \lgs spots.  We find (Fig.~\ref{fig:lgsncen}(a))
that is only slightly affected by sub-aperture size, until \lgs spots
become more severely truncated (Fig.~\ref{fig:lgsncen}(b)).

\subsection{Exploration of LGS number}
\begin{figure}
  \includegraphics[width=\linewidth]{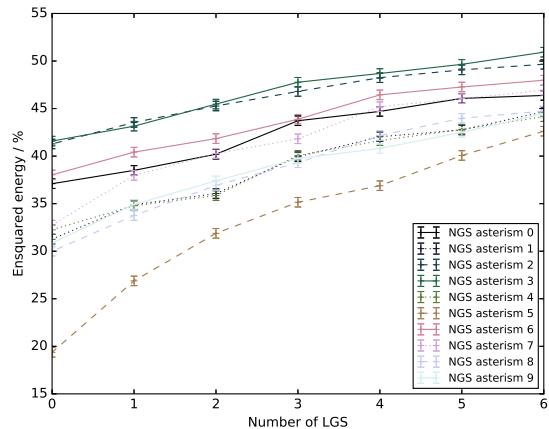}
  \caption{A figure showing on-axis \ao performance (H-band ensquared energy
    within a 150~mas box) for different numbers of LGSs, and different
    NGS asterisms.}
  \label{fig:nlgs}
\end{figure}

We have previously explored \moao performance with a number of
different \ngs asterisms \citep{basden17}.  Here, for each of these
asterisms, we investigate performance as a function of number of \lgss
used, as shown in Fig.~\ref{fig:nlgs}.  It can be seen that the rate
of drop in performance is dependent on the \ngs asterism, and that in
particular, asterism 5 gives poorer performance.  We note that
this particular asterism contains only 2 stars, and that one of these
is very faint (2.9 photons per pixel per frame at 250~Hz).

We note that this result does not point to a conclusive \lgs
requirement.  Performance is seen to increase as guide star number
increases, and for particular \ngs asterisms can drop off sharply when
few \lgss are used.  It is encouraging that good performance is seen
with 4 \lgss, since this is the current baseline number that will be
provided by \eso at the \eelt.  Whilst maximum performance for the
best asterisms drops from 52\% to 48\% as \lgs number is decreased
from 6 to 4, performance with the worst performing asterism drops from
42\% to 37\%.

We also note that when number of \lgs is reduced, it would be possible
to compensate the performance loss by increasing the number of \ngss.
However, this is not something that we investigate, partly due to the
reduction in sky-coverage that would ensue, and because of the
increase in system complexity with greater numbers of natural guide
star acquisition systems.

\subsection{Reduction in frame rate and sub-aperture count of faint NGSs}
\begin{figure}
  \includegraphics[width=\linewidth]{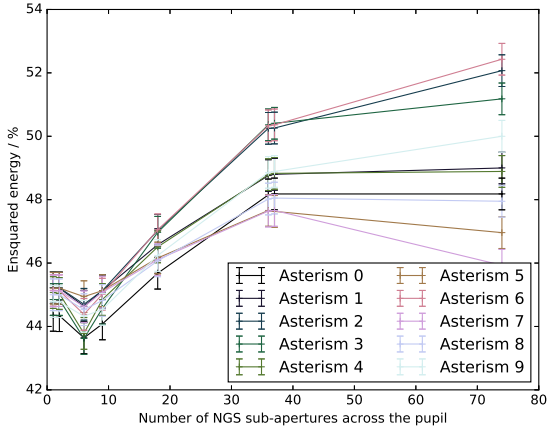}
  \caption{A figure showing on-axis \ao performance (H-band ensquared energy
    within a 150~mas box) as a function of number of NGS
    sub-apertures, for the different NGS asterisms under consideration.}
  \label{fig:nsubx}
\end{figure}

Given that some of the \ngss within our chosen asterisms are extremely
faint, there are two obvious ways in which flux can be increased.
Either, sub-aperture count can be reduced (so spreading available flux
between fewer sub-apertures), or the \wfs frame rate can be reduced
(allowing longer integration times).  Having a variable sub-aperture
count in an \ao system is not always simple if Shack-Hartmann systems
are used: an optical realignment is required, and mechanisms are
required to change the lenslet arrays.  This is not an attractive
prospect for an \elt-scale \moao system, however one possibility may
be to have a pre-defined number of high and low order wavefront
sensors present, which can then be selected depending on target
availability.  Alternatively, a Pyramid wavefront sensor
\citep{pyramid} can be used, which provides the ability to re-bin
wavefront sensor images on-the-fly.  Here, we do not consider the use
of Pyramid sensors as the relative merits of Shack-Hartmann and
Pyramid systems have been explored elsewhere, and are not part of the
baseline design for the current \eelt \moao instrument concept.
Additionally, pyramid sensors are non-linear, which for a partially
open-loop instrument such as the \moao instruments simulated here,
introduces significant additional complexity.  Techniques to combine
pyramid and Shack-Hartmann wavefront sensor signals in a tomographic
\ao system are also not well studied.  However we note that this would
potentially be another way to improve performance.  To simplify our
investigation, we consider the case in which the sub-aperture count of
all \ngs \wfss is reduced, and expected performance is shown in
Fig.~\ref{fig:nsubx}.  It can be seen that a factor of two reduction
in sub-aperture count (across the pupil) can lead to slight
performance improvement for some asterisms under consideration, but
that further reductions lead to worse performance, i.e.\ \ngs
information is required for tomographic wavefront reconstruction.
However, we suggest that in cases where sky coverage is important,
\ngs \wfs sub-aperture count could be reduced, increasing coverage,
but leading to slightly lower performance overall.  We also consider
the case where slope measurements from a high order ($74\times74$)
sensor are averaged to give a global tip-tilt measurement (i.e. using
the \ngs for tip-tilt only, without requiring optical modifications).
However, this yields a more significant drop in performance (with
H-band ensquared energy within 150~mas dropping to about 30\%).
Therefore, we do not consider this option further here.  \ignore{For
  comparative purposes, we note that performance of a \lgs-only system
  (assuming a valid \lgs tip-tilt signal, which is of course
  nonphysical) gives ensquared energy within a 150~mas box of about
  24\%.}

\begin{figure}
  \includegraphics[width=\linewidth]{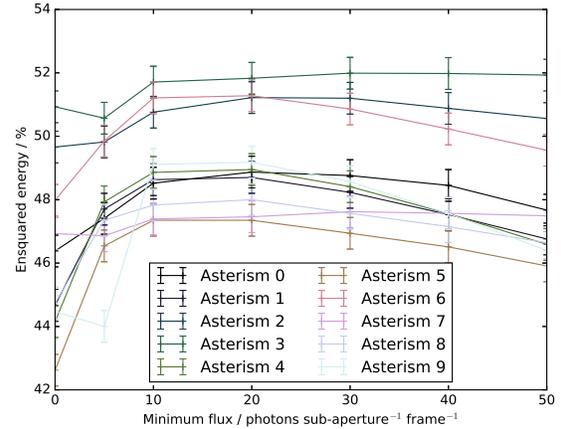}
  \includegraphics[width=\linewidth]{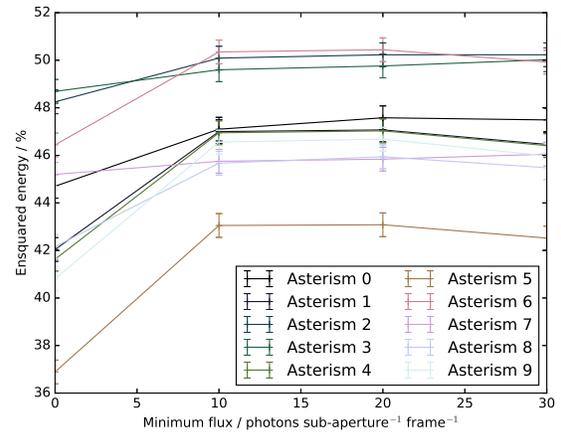}
  \caption{A figure showing on-axis \ao performance (H-band ensquared energy
    within a 150~mas box) as NGS frame rate is reduced so that a given
    minimum flux is recorded (given by the x-axis), for (a) 6 LGS, and (b) 4 LGS.}
  \label{fig:framerate}
\end{figure}

A reduction in faint \ngs \wfs frame rate can also be used to increase
detected \ngs flux, and one which can be performed entirely by the software
controlling the \ao system (i.e.\ no changes to the optical or
mechanical design are required).  To investigate this, we take a
pragmatic approach: we assume that the \lgs \wfss and bright \ngs
\wfss will operate at the baseline frame rate of 250~Hz.  We then
reduce the frame rate of fainter \ngs \wfss in steps of size equal to
the time period of the \lgs \wfss (4~ms), until the detected flux for
these \wfss reaches some set level.  For example, if we specify a
minimum flux of 20 photons per sub-aperture per frame, then a \wfs
that measures 15 photons at 250~Hz would be reduced to 125~Hz
(delivering 30 photons) and a \wfs that measures 7 photons at 250~Hz
would be reduced to 83~Hz (3 time steps), delivering 21 photons.
Fig.~\ref{fig:framerate} shows predicted on-axis \ao performance for
this investigation.  It is clear that this approach can significantly
increase \ao performance when sources are faint.  In general, we find
that a minimum flux of about 10 detected photons per sub-aperture offers best
performance for the asterisms studied.  

\subsection{Investigation of commercial WFS detector configurations}
\label{sect:ixon}
In order to reduce risk associated with development of a \moao system,
we here consider the use of a commercially available detector for the
\ngs \wfss, a $1024\times1024$ \emccd.  We use camera specifications
of an Andor Technologies iXon Ultra 888 \citep{ixonUltra888} which has
a maximum frame rate (full frame) of 26~Hz.  We also note than an
Imperx Puma camera \citep{imperx} would also meet these
specifications.  We assume 0.1 electrons
readout noise, and the modes of operation that we consider are given
in table~\ref{tab:cameras}.  \ignore{We consider the case with standard \ngs
flux as given in Fig.~\ref{fig:asterism}, and also with half the flux,
to simulate the \emccd excess noise factor.}  We have focused on this
detector for several reasons:
\begin{enumerate}
\item It has enough pixels to provide a reasonable sub-aperture size,
  to avoid centroid gain variations due to changes in seeing.
  \item It has low readout noise, essential for increasing sky
    coverage.
    \item It has a frame rate that (as we show) is sufficient to not
      significantly affect \ao performance.
\end{enumerate}

For the \ngs \wfss, we also consider the use of the proposed (not yet
available) \eso LGSD dectector with a $1760\times1680$ pixels, and a
maximum frame rate of over 250~Hz \citep{lgsdshort}.  This detector is
specified with a readout noise of 3 electrons.

\begin{table}
  \begin{tabularx}{\linewidth}{lllll}\\ \hline
    Window size & AFR (MFR)  & Pixels per & Number of LGS\\
    &/ Hz&sub-aperture & integrations per NGS \\ &&&frame \\ \hline
$962\times962$ & 27.8 (30) & $13\times13$ & 9 \\
$888\times888$ & 35.7 (36) & $12\times12$ & 7 \\
$740\times740$ & 50 (52) & $10\times10$ & 5 \\
$592\times592$ & 62.5 (75) & $8\times8$   & 4 \\
    $444\times444$ & 83.3 (110) & $6\times6$  & 3 \\
  \end{tabularx}
  \caption{A table showing possible modes of operation for the NGS using a
    $1024\times1024$ pixel EMCCD.  The maximum frame rate (MFR) of the
  camera for each mode of operation is given, though we reduce this
  for each case to ensure a whole number of 250~Hz LGS integrations
  fit in each NGS frame period, given by the actual frame rate (AFR).}
  \label{tab:cameras}
\end{table}

For the \lgs \wfss we assume $16\times16$ pixels per sub-aperture and
3 electrons readout noise.  We note that there are two distinct
detector possibilities that can meet this specification, though
neither is yet commercially available in a suitable camera.  The first
is the \eso LGSD \citep{lgsdshort} (manufactured by E2V), while the
second is the Fairchild Imaging LTN4625A \scmos detector \citep{ltn4625a}.

\begin{figure}
  \includegraphics[width=\linewidth]{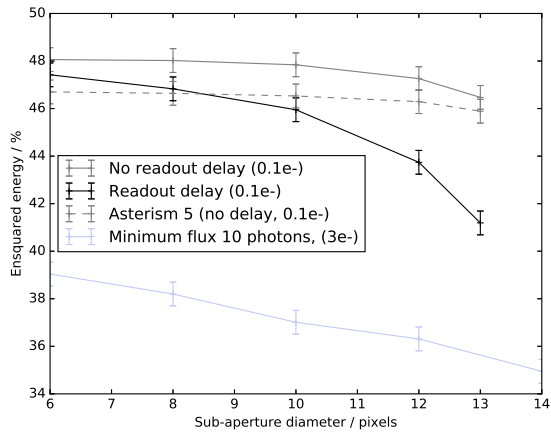}
  \caption{A figure showing \ao performance (on-axis H-band ensquared energy
    in a 150~mas box) for the different \ngs \wfs camera options described
    in the text and legend.  Asterism 0 is used unless otherwise
    stated.  For the 0.1e- readout noise case, frame rate is reduced
    as per table~\ref{tab:cameras}, while a frame rate of 250~Hz is assumed for the
    3e- readout noise case, reduced for faint guide stars such that at
    least 10
    photons per sub-aperture per frame are delivered.}
  \label{fig:camera}
  \end{figure}

For the \ngs \wfss, we investigate the trade-off between maximum frame
rate, and number of pixels per sub-aperture in Fig.~\ref{fig:camera}.
Here we can see that better performance is achieved using a higher
frame rate, and hence fewer pixels per sub-aperture.  For comparison,
the case using the LGSD detector for the \ngs \wfss is also shown, and it is
evident that performance is far worse, primarily due to the higher
readout noise.  We therefore recommend the use of lower readout noise
detectors for the \ngs \wfss even if these are required to run at
lower frame rates.

When the detector with 0.1 electrons readout noise is used, frame rate
must be reduced (Table~\ref{tab:cameras}) due to camera readout
modes.  Fig.~\ref{fig:camera} displays two sets of information
for clarity.  Firstly, we reduce the camera frame rate (increase the
exposure time), but do not add any additional readout delay (signified
by ``no readout delay'' in the legend).  We note
that is unphysical with this detector.  We therefore also reduce the frame rate, and
increase the readout time, to give actual achieved performance
(signified by ``Readout delay'' in the legend).  It is
therefore evident that this additional delay begins to have
significant impact, and that therefore, smaller sub-apertures (with
increased frame rate) are favoured.  For comparison, we also show
performance using \ngs asterism 5, demonstrating that good performance
can be achieved even in the challenging case of the sparsest
identified \ngs asterism.

\ignore{
  Run: Also minFlux for the 3e- readnoise case.
xxx
TO RUN: with 0.5x flux.

6       3       3.0     0.1     4.127   48.06
8       4       3.0     0.1     4.787   48.02
10      5       3.0     0.1     4.991   47.84
12      7       3.0     0.1     5.033   47.26
13      9       3.0     0.1     4.877   46.47

Ast 5:
6       3       3.0     0.1     5       6.231   46.7
8       4       3.0     0.1     5       6.487   46.64
10      5       3.0     0.1     5       6.564   46.53
12      7       3.0     0.1     5       6.561   46.29
13      9       3.0     0.1     5       6.468   45.89

additional ixonDelay:

6       3       3.0     0.1     0       10      3       3.962   47.42
6       3       3.0     0.1     0       20      3       3.956   47.34
6       3       3.0     0.1     0       2       3       3.957   47.42
8       4       3.0     0.1     0       10      4       4.461   46.83
8       4       3.0     0.1     0       2       4       4.453   46.83

Using 3e- for NGS:
5000.0	0.4	3.0	12.0	0.5576	37.38

todo: Run some minflux cases for this - 5, 10, 20, 30.
48	5	0.3249	32.76
48	10	0.3458	33.16
48	20	0.4243	34.94
48	30	0.00367	0.8385
Hmm - not so good - also run some with noisePower=0.4.

Base case for comparison: 3e- for lgs, 0.1 for ngs, 250Hz ngs, 16x16.
16      1       3.0     0.1     0       2       None    4.216   46.91
}

We note that we do not include the impact of telescope vibrations
within these results.  Therefore the results given at lower frame
rates are likely to be more optimistic, since lower frame rates will
suffer more from uncorrected vibrations, even when vibration
mitigation techniques, such as \lqg control, are used
\citep{lqgOptExpshort}.  Higher order vibrational modes can be
compensated using \lgs measurements.  It is only the lowest order
modes (tip-tilt), with correspondingly lower frequencies, which would
require \ngs signals for correction.

Our recommendation for wavefront sensor detector technology is
therefore to use low readout noise \emccd detectors for the \ngs \wfss,
which can be operated at a lower frame rate, and higher readout noise
\cmos detectors for the \lgss, where larger detector area is important
to prevent performance degradation due to spot truncation, and where
incident flux can be higher such that the increased readout noise has
less of an impact.  The increased readout rates of large \cmos devices
(compared with equivalent sized \emccds) is also advantageous for the
\lgss.

\subsection{NGS flux investigation}
\begin{figure}
  \includegraphics[width=\linewidth]{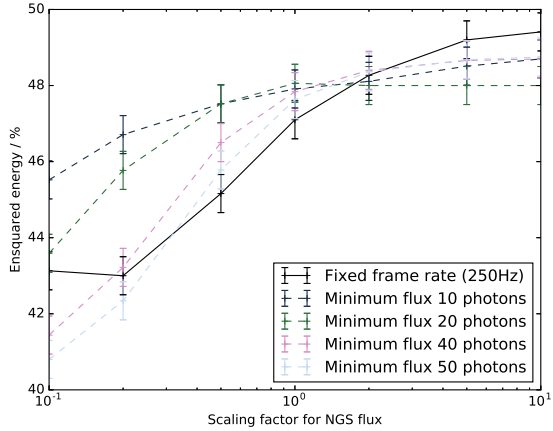}
  \caption{A figure showing \ao performance (H-band ensquared energy
    within 150~mas) as a function of globally scaled \ngs flux.  Shown
    are plots for all WFSs having the same frame rate (250~Hz),
    and where NGS WFSs have reduced frame rate such that a minimum
    flux level is recorded, as given in the legend (in photons per
    sub-aperture per frame).}
  \label{fig:ngsfluxscaling}
\end{figure}

Since instrumental and telescope optical throughputs are not known, we
investigate performance as a function of \ngs flux, i.e.\ we apply a
global scaling to the flux provided by a given \ngs asterism.
Fig.~\ref{fig:ngsfluxscaling} shows these results for one \ngs
asterism (asterism 0), and these can be
compared with \citet{basden17}.  This figure also demonstrates how \ao
performance can be improved by reducing the \wfs frame rate used with low-flux
\ngss.  Here, we can see that in the case of the \ngs asterism 0 studied
here, by reducing \ngs \wfs frame rates so that each sub-aperture contains
at least 10 photons per frame, an improvement in performance can be
seen compared with when all wavefront sensors operate at the same
frame rate.  For comparison, when no \ngss are used (and the \lgs
tip-tilt signal is assumed valid), ensquared energy within a 150~mas
box is found to be about 24\%.

\subsection{LGS flux investigation}
In contrary to \ngs flux which is known, \lgs flux will vary depending
on conditions within the sodium layer.  We therefore explore \ao
performance as a function of \lgs return flux level.  We consider
several different detector scenarios, each with different readout
noise levels.  First, we consider the case where the \lgs and \ngs \wfs
cameras have identical readout noise (from 0.1 to 3 electrons).  We
then also consider the case where the \ngs \wfs readout noise is fixed at
0.1 electrons (since we know that such a detector exists, as described
in \S\ref{sect:ixon}), while \lgs readout noise varies.  The range of \lgs
flux used is given from measured on-sky flux return \citep{caliaPrivate}.  

\begin{figure}
  \includegraphics[width=\linewidth]{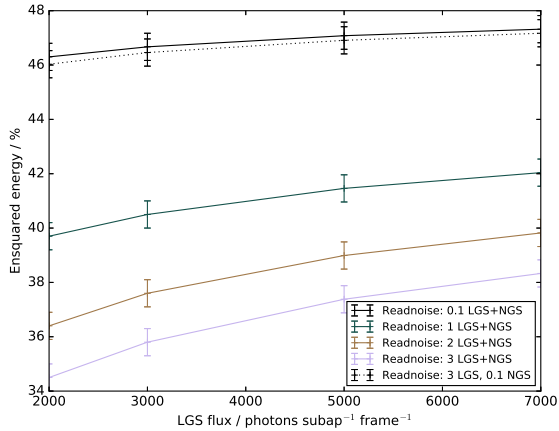}
  \caption{A figure showing \ao performance (H-band ensquared energy
    within 150~mas) as a function of LGS flux (in photons per
    sub-aperture per frame), for different detector readout noise
    levels, as given in the legend.  For solid curves, the LGS and NGS
    detectors have the same readout noise (given in the legend).  For
    the dashed curve, the NGS has a readout noise of 0.1 electrons,
    while the LGS readout noise is given in the legend.}
  \label{fig:lgsflux}
\end{figure}

Fig.~\ref{fig:lgsflux} shows the results, and it can be seen that when
the \ngs \wfs readout noise is low, the expected \lgs flux is sufficient to maintain
good \ao performance.  However, when both \lgs and \ngs \wfs readout noise
increase \ao performance drops.  Therefore, if instrumental trade-offs
must be made related to detector readout noise, it would be
advantageous to ensure that low readout noise for the \ngs \wfs is
prioritised.  \lgs \wfs readout noise level is not critical with the
expected \lgs flux return at these frame rates.

\subsection{LGS asterism diameter}
\begin{figure}
  \includegraphics[width=\linewidth]{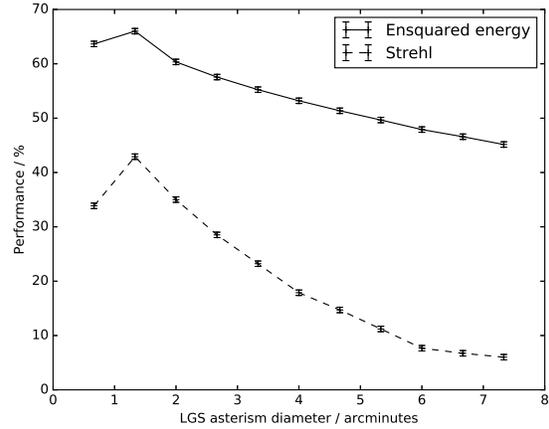}
  \caption{A figure showing \ao performance at H-band as a function of LGS
    asterism diameter.  Ensquared energy is within a 150~mas box
    size.}
  \label{fig:lgsdiam}
\end{figure}

Although the performance of \ltao and \moao as a function of \lgs
asterism diameter has been well studied elsewhere
\citep{miskaltao,basden12}, we include here such a study, to provide a
reference, and for comparisons with other results.
Fig.~\ref{fig:lgsdiam} provides this information for both H-band
Strehl ratio and ensquared energy (with a 150~mas box size).

\subsection{Guide star asterism rotation}
When operating multiple \lgs \ao systems, there are two possibilities
for \lgs tracking: either they can track the telescope pupil, or track
the sky rotation during observations.  In the former case, relative
alignment between the \lgs and un-de-rotated components (such as M4)
will remain constant, while relative alignment between the \lgs and
\ngs \wfss will change. Therefore, continual update of the tomographic
reconstruction matrix will be necessary to account for the change in
relative \wfs alignment.  The required frequency of this update is
determined by the tolerance of \ao performance to mis-rotation.

In the latter case, relative alignment between M4
and the \wfss will change, and it is necessary to be able to steer the
\lgss (to maintain tracking).  The relative \wfs alignments will also
change with flexure and rotation of lenslet arrays with respect to
other guide stars.

\ignore{
However, \wfs alignment will remain
constant, and so tomographic reconstructor update will not be
necessary, rather just a rotation during the \dm fitting step.
}

\begin{figure}
  \includegraphics[width=\linewidth]{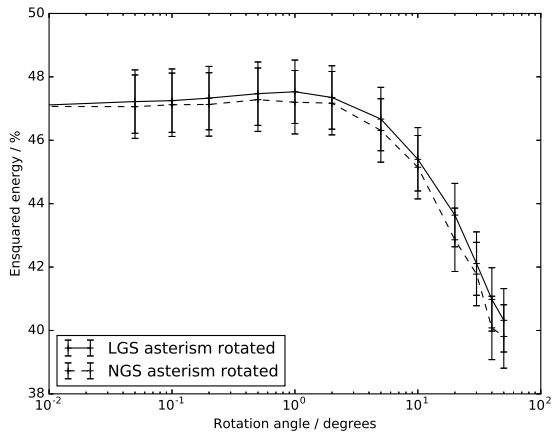}
  \caption{A figure showing \ao performance (H-band ensquared energy
    within a 150~mas box) as a function of differential rotation
    between \dms and on-sky guide star position.}
  \label{fig:rotation}
\end{figure}

The impact of sub-aperture rotation has been studied previously
\citep{basden12}.  Here, we consider only the effect of \lgs motion
(tracking the sky) resulting in a change in tomography, i.e.\ in
portion of the atmosphere through which the \lgs propagates.  These
simulations do not include the effect of rotation of wavefront sensors
with respect to others.  Fig.~\ref{fig:rotation} shows \ao performance
as a function of rotation angle between the on-sky guide star
positions, and provides information on how frequently a wavefront
reconstruction matrix must be updated to maintain performance during
\ao operation.  Here, we rotate the on-sky position of either the
\lgss or the \ngss.  Sub-aperture alignment remains constant,
i.e.\ aligned with \dm actuators.  This figure therefore shows how
performance changes as the guide stars sample different parts of the
turbulent volume than the tomographic reconstruction expects.  We note
that the drop in performance is relatively small.  This is to be
expected because ground layer sampling remains unaffected (it is
sampled regardless of where the guide stars point), and the ground
layer encompasses a significant amount of turbulent
strength. \ignore{and also because sensing and correction of
  spherically symmetric modes will also be unaffected (e.g.\ focus and
  coma).}  We see that performance is maintained for differential
rotations of up to about one degree.  Therefore, update of the
tomographic control matrix is necessary whenever the relative
sky-pupil rotation becomes larger than one degree.  We note that this
figure does not include any differential rotation of the \wfs
sub-aperture alignment, or between the \dm actuator grids and the
sub-aperture grids.  These effects could have a large impact in
practice, and so in the case where the wavefront sensor alignment is
not maintained (by optical derotation), further investigation would be
necessary.  

\subsection{DM actuator count}
\dms are key components within an \ao system, and for a \moao system
which requires one \dm per channel, can introduce a significant cost
to the overall design.  Higher order \dms are generally more
expensive, and therefore it may be attractive to reduce system cost by
reducing the number of \dm actuators.  We have therefore investigated
\ao performance when using \moao \dms of different orders, as shown in
table~\ref{tab:dmorder}, for 3 currently available \dm sizes.  We see
that reducing \dm order leads to a drop in \ao performance.  However,
individual instrumental science requirements may mean that performance
goals can still be met with a lower performance, and so using lower
order \dms should not be ruled out simply because performance is
reduced.  We note that these results are similar to those reported by
\citet{basden12}.  

\begin{table}
  \begin{tabular}{llll}\\ \hline
    M4 actuator geometry & MOAO  & Strehl & EE\\
    & actuator count && \\ \hline
Square $75\times75$ &$64\times64$&	6.77&	46.9\\
Square $75\times75$ &$32\times32$&      4.38&    41.0\\
Square $75\times75$ &$17\times17$&      2.88&    35.7\\
Hexagonal $75\times65$ &$64\times64$& 6.31&	46.0\\ \hline
  \end{tabular}
\caption{A table showing H-band on-axis AO performance for different DM
  configurations (EE is ensquared energy is within a 150~mas box size).}
\label{tab:dmorder}
\end{table}

We also consider the use of a hexagonal actuator pattern for M4, (the
global ground layer correction), with the \dm having $75\times65$
actuators.  As expected, this only has a small impact on performance
(given in the table), and so, for generality, we have used a square actuator
pattern throughout this paper.

\subsection{Sensitivity to atmospheric turbulence profiles}
We have been using median seeing from the standard \eso 35-layer
atmospheric profile \citep{35layer} for these simulations.  However,
under different seeing conditions, \ao performance can be
significantly different.  Table~\ref{tab:seeing} shows how \ao
performance is expected to vary using the defined 4 quartile
profiles.  Additionally, \ao performance when using an alternative
31-layer atmospheric profile (which is defined in \eso internal
document ESO-191766v7) is also given.  We see
here that under poor atmospheric conditions, \ao performance is
significantly worse.  Therefore when designing an \ao instrument to
meet specific science requirements, it is important to specify
observing condition, i.e.\ should science requirements always be
achievable, or only some fraction of the time.

\begin{table}
  \begin{tabularx}{\linewidth}{llll}\\ \hline
    Profile & Strehl & Ensquared energy & GLAO ensquared\\
    &&&energy\\ \hline

Quartile 1	&17.3&	61.2 & 42.7\\
Quartile 2  &     8.4 &48.3 & 29.9\\
Median & 5.2 &	46.9 & 24.0 \\
Quartile 3       &3.0 &36.9 & 19.5\\
Quartile 4	&0.7	&22.1 & 5.9\\
31-layer &5.6     &47.4 & 12.6\\ \hline

  \end{tabularx}
  \caption{A table showing H-band \ao performance under different atmospheric
    conditions. Ensquared energy is within a 150~mas box.}
  \label{tab:seeing}
\end{table}

\subsection{GLAO performance}
An \moao system is able to perform tomographic \glao correction for
free, using either a common \dm (e.g.\ \eelt M4), or the individual
\moao \dms (since these are usually ground conjugated).

\begin{figure}
  \includegraphics[width=\linewidth]{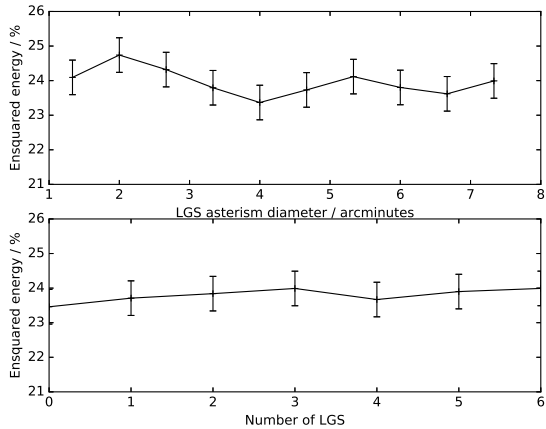}
  \caption{A figure showing GLAO performance (H-band ensquared energy
    within a 150~mas box) as a function of LGS asterism diameter (with
    6 LGS), and number of LGS (with a 7.3~arcminute asterism
    diameter).}
  \label{fig:glao}
\end{figure}

We therefore investigate \glao performance as a function of \lgs
asterism diameter, and number of \lgss.  As seen in
Fig.~\ref{fig:glao}, performance is fairly insensitive to these
changes, since all configurations are able to identify the ground
layer.  However, we note that the correction quality is significantly
degraded from that achieved using \moao, i.e. \glao ensquared energy
is only about half of the \moao energy for the given atmospheric
profile.  Here, we have assumed that the \glao \dm is conjugated at
zero.  However, as mentioned previously, the \eelt M4 \dm is expected
to be conjugated at about 625~m.  A previous study \citep{basden21}
has shown that this difference is expected to have little impact on
performance.

\subsection{Comparisons with other simulation results}
Direct comparison with previous results from other simulation tools is
not trivial, due to significant differences in input parameters
(wavelengths, atmosphere models, sub-aperture count, etc.)  However, a
study of performance trends is possible, and we find that several of
the performance trends that we present here (e.g.\ performance as a
function of asterism diameter, with scaling of \ngs flux, number of
\lgs, and \lgs pixel scale) are similar to those in
previous studies using Monte-Carlo end-to-end \ao simulation 
\citep{basden21,miskaltao,2011aoel.confE..63T,2010aoel.confE2013F,basden17,basden12}.
We note that Monte-Carlo models are pessimistic when compared to
analytical model results \citep{2008JOSAA..26..219N}.

\section{Conclusions}
We have performed detailed Monte-Carlo modelling of several of the
design parameters for a potential \eelt \moao instrument using a full
Monte-Carlo end-to-end \ao simulation tool, \dasp.  The
recommendations that we draw from this study include \lgs pixel scale
(typically optimal at 0.7~arcseconds per pixel),
minimum sub-aperture size (at least $10\times10$ pixels), a study of
number of guide stars, and reduction in \ngs \wfs frame rates so that a
minimum detected flux is received (at least 10 photons per
sub-aperture when using an \emccd).  We identify current commercial
cameras that would be suitable for wavefront sensors (reducing the
risk associated with such an instrument).  We include a study of
several demanding \ngs asterisms, taken from availability of stars
within a cosmological field, and also consider performance as a
function of \lgs return flux.  We also consider different \dm sizes
and geometries and the effect of differential rotation between \dms
and on-sky guide star positions.  Although Strehl ratios are typically
fairly low (6--10\% at H-band), the ensquared energy requirements for
a typical spectrograph (e.g.\ MOSAIC) are likely to be met, being in
the range of 50\% for energy within a 150~mas box.  

\section*{Acknowledgements}
This work is funded by the UK Science and Technology Facilities
Council, grant ST/K003569/1, and a consolidated grant ST/L00075X/1.

\bibliographystyle{mn2e}

\bibliography{mybib}
\bsp

\end{document}